\documentclass[aps,prd,superscriptaddress,floats]{revtex4}

\begin{document}
\title{Vortices in Theories with Flat Directions}
\author{A. Ach\'{u}carro}
\affiliation{Lorentz Institute of Theoretical Physics, University of Leiden, Leiden, The Netherlands}
\affiliation{Department of Theoretical Physics, UPV-EHU, Bilbao, Spain}
\affiliation{Institute for Theoretical Physics, University of Groningen, Groningen, The
Netherlands}
\author{A.C. Davis}
\affiliation{DAMPT, Centre for Mathematical Sciences, Cambridge University, Cambridge, U.K.}
\author{M. Pickles}
\affiliation{DAMPT, Centre for Mathematical Sciences, Cambridge University, Cambridge, U.K.}
\author{J. Urrestilla}
\affiliation{Department of Theoretical Physics, UPV-EHU, Bilbao, Spain}
\date{\today}
\begin{abstract}

In theories with flat directions containing vortices, such as
supersymmetric QED, there is a vacuum selection effect in the allowed
asymptotic configurations. We explain the role played by gauge fields
in this effect and give a simple criterion for determining what vacua
will be chosen, namely those that minimise the vector mass.  We then
consider the effect of vacuum selection on stable (BPS)
non--topological vortices in a simple Abelian model with $N\!=\!2$
supersymmetry which occurs as a low energy limit of Calabi--Yau
compactifications of type II superstrings.  In this case the magnetic
flux spreads over an arbitrarily large area.  We discuss the
implications for cosmology and for superstring inspired magnetic
confinement scenarios.

\end{abstract}

\maketitle

\section{Introduction}
Flat directions of the scalar potential are a feature of a large class of
supersymmetric (SUSY) theories, for instance in Abelian theories
where the gauge symmetry is broken with a Fayet-Iliopoulos (FI) D-term. 
In the last few years these have undergone intense study, in particular 
in the context of effective actions for SUSY non-Abelian gauge theories
and confinement (see e.g. \cite{W.,dterm}).

A generic consequence of SUSY theories with a D-term is the formation of
cosmic strings \cite{PRTT96,DDT97}. However, the string solution is
richer than the usual Nielsen--Olesen (NO) vortex \cite{A57,NO73} for
two distinct reasons. Firstly, the existence of flat directions means
that there is a moduli of degenerate vacua and, at first sight, it is
unclear which of the vacuum states give rise to stable cosmic string
solutions. Secondly, the presence of supersymmetry gives rise to
fermion zero modes in the string core \cite{DDT97}, rendering the
string current--carrying. This typically leads to the formation of
{\it vortons}, long--lived remnants whose observational signatures can
severely constrain, and in some cases rule out, such models\cite{chiralvorton}.

Topological strings arising from the bosonic sector of an $N\!=\!1$
 SUSY model have been analysed in \cite{PRTT96}, showing that even
 though there is a one-parameter choice of vacuum for the moduli
 field, only one choice permits cosmic string solutions which could be
 stable. Their conclusion was based on the behaviour of the modulus
 field well outside the core of the vortex. This vacuum selection
 effect is generic in Abelian theories with flat directions in that,
 as we will show, it is driven by the tendency to minimise the vector
 mass at the core of the vortex. Thus, one expects it to affect
 non--topological as well as topological strings, and  it could
 conceivably improve their stability.

In this paper we analyse this vacuum selection effect in the context
of (BPS) non--topological strings.  In particular, we investigate
the nature of BPS vortices in $N\!=\!2$ supersymmetric QED with two
hypermultiplets of opposite charge. This model has been analysed
previously in connection with Calabi--Yau compactifications of type II
superstrings \cite{GMV96}, where the existence of vortices is linked to the
confinement of magnetic charges. We will find that, due to the
vacuum selection effect, the structure of the BPS vortices is
identical to that of so-called semilocal strings\cite{VA91}.

Apart from its physical significance, this model has the 
advantage that all calculations can be performed explicitly, which is
important because the stability of semilocal strings is somewhat
counter-intuitive.  We should stress that although the vacuum manifold
is simply connected, the gauge boson is massive and magnetic flux is
topologically conserved and quantized; the vortices analysed here
are {\it bona-fide} BPS states carrying one unit of magnetic flux.

And yet we will find that, {\it in practice, there are no stable
vortices in this model}. To be more precise, the vortices are only
neutrally stable (which is consistent with the BPS condition) and
degenerate with a whole family of other BPS ``magnetic flux tubes'' of
arbitrarily large width. The slightest perturbation excites the zero
mode and makes the vortex expand.  The vacuum selection effect does
not single out the narrow Nielsen--Olesen vortex over the other, more
extended, BPS flux tubes.  In a cosmological context, this suppresses
vorton production. In the context of superstring compactifications, it
casts serious doubts on the mechanism of magnetic charge confinement
proposed in \cite{GMV96}.

The paper is organized as follows. We first review the vacuum
selection effect in the model of \cite{PRTT96}.  We explicitly show
that the resulting cosmic string saturates the Bogomol'nyi bound, so
is strictly stable, and that it is no accident that the selected
vacuum is the one minimising the vector boson mass. We argue that this
gives a generic criterion to determine which vacuum or vacua are
selected in a given theory, provided all charges are equal in absolute
value.

We then extend this idea to $N\!=\!2$ QED with two hypermultiplets of
opposite charge, using a Bogomol'nyi argument on the corresponding
solution \cite{edel}. 
We again show that it saturates a
Bogomol'nyi bound, so is stable, but the presence of a cylindrically
symmetric zero mode changes the dynamics completely and in particular
the string expands to an arbitrarily large area in finite time
\cite{H92,leese,AV00}. This occurs even though the vector mass is
non-zero and the magnetic flux at infinity remains quantised, setting
a constraint on viable models of confinement of magnetic charge in
these N=2 Abelian theories. Finally we discuss the effect of SUSY breaking on the solutions and
speculate on the resulting cosmology.

\section{Vacuum selection effect and minimum gauge mass}

We begin by reviewing the vacuum selection effect described in  
\cite{PRTT96}. Penin, Rubakov, Tinyakov and Troitsky considered a
four-dimensional model with two $N\!=\!1$ chiral superfields of
opposite charges coupled to a U(1) vector multiplet, and a
FI  D-term which gives the symmetry breaking potential. The action of
the bosonic sector is
\begin{eqnarray}
&S =\!\! \int d^4x\left[|D_\mu \phi_+|^2 + |D_\mu \phi_-|^2- \frac14{F_{\mu\nu} F^{\mu\nu}} 
- V(\phi_+,\phi_-)\right]&\nonumber\\
&V(\phi_+,\phi_-)={\frac\lambda 2}(|\phi_+|^2 - |\phi_-|^2 - \eta^2)^2&\label{v}
\end{eqnarray}
where $A_\mu$ is a $U(1)$ gauge field 
and $\phi_\pm$ are 
complex scalars of opposite charges ($q_{\pm}=\pm q$),
\[
D_{\mu}\phi_{\pm} = (\partial_{\mu} \pm i\, q\, A_{\mu})\,\phi_\pm, \qquad
F_{\mu\nu} = \partial_\mu A_\nu - \partial_\nu A_\mu.
\]
The solutions for the vacuum manifold ($V=0$) are 
\begin{equation}
|\phi_+| = \eta \cosh u \equiv v_+, \ |\phi_-| = \eta \sinh u \equiv v_- \ ,
\label{vacman}
\end{equation}
where  $u$ parametrises the moduli space. After symmetry breaking the
physical spectrum consists of two massless scalars, the Goldstone boson 
and the modulus field; a massive scalar particle $m_s^2=2\lambda\eta^2
\cosh 2u$; and a massive vector particle $m_v^{\ 2}= 2 q^2 \eta^2 \cosh 2u$.
Their masses depend on the choice of vacuum.  
The minimum values ${\bar m_s}$ and 
${\bar m_v}$ correspond to $u=0$ ($\phi_-=0$), where we recover the Abelian
 Higgs model. This has static vortex solutions, the NO strings \cite{NO73}. 

We want straight, static vortices along the z-direction so we drop the
t- and z-dependence and set~$A_t$~$=$~$A_z$~$=$~$0$, defining
$B=\partial_1A_2-\partial_2A_1$. We can set $\eta=q=1$ by rescaling,
and we consider the Bogomol'nyi limit $\lambda=1$ {this effectively
makes the model N=2 supersymmetric with one hypermultiplet, see also
\cite{edel,FG98,hou}}.  The conditions for finite energy include
$D_\mu \phi_+$~$\to$~$0,$ $\ D_\mu \phi_- \to 0 , B \to 0 $ faster
than $1/r$ and this introduces a correlation in the phases of the
fields at infinity: fields with charge $q_a$ wind as $e^{inq_a
\theta}$ and the gauge field tends to a constant, $A_\theta \to -n$,
leading to the quantisation of magnetic flux, $\int d^2 B = -2\pi n$.

In principle, one can try to construct vortices tending to any of the
vacua (\ref{vacman}) as $r \to \infty$.  However, in \cite{PRTT96} it
was shown that the only static solutions are those with $u=0$.  Any
other choice of boundary conditions leads to what is effectively an
unstable vortex that tends to this one. The instability is very mild,
and the surviving string is a NO string. We will now show that this
string is stable.

In what follows we consider cylindrically symmetric configurations
with $n=1$ :
\[
\phi_+ = f_+(r) e^{i\theta} \qquad
\phi_- = f_-(r) e^{-i\theta}e^{i\Delta} \qquad
A_\theta  d\theta= a(r) d\theta 
\]
$f_\pm(r)$ are real functions, with  
$f_+(0) = f_-(0)=a(0)=0$ and, 
$f_\pm(\infty)=v_\pm$, $a(\infty)=-1$.
 $r$--dependent phases $e^{i\psi_\pm(r)}$
for $\phi_\pm$ are possible but 
minimum energy requires 
$\partial_r \psi_\pm = 0$, so we will ignore them. 
$\Delta$ is a real constant.

To prove stability we obtain the Bogomol'nyi equations
\begin{equation}
f'_\pm \mp \frac{a+1}r f_\pm = 0, \qquad {a'\over r} - (f_+^2 - f_-^2 -1) = 0. \label{aeq} 
\end{equation}
Note that, if $f_-(r)=0$, eqns. (\ref{aeq}) are 
the standard Bogomol'nyi  equations for the Abelian Higgs model.

It can be seen that the only solution for $f_-$ satisfying the boundary
conditions at the core is $f_-(r)=0$; thus the NO solution saturates
the Bogomol'nyi bound automatically. This shows that the string is
stable, since it is a  global minimum of the energy, and excludes the
possibility of cylindrically symmetric zero modes. In fact it turns
out that the restriction  to cylindrical symmetry is unnecessary:
$\phi_-$ always vanishes on the solutions to the Bogomol'nyi equations
(see, e.g. \cite{W.}).  The only remaining solutions are the standard
NO vortices.

As explained in \cite{PRTT96}, the key point in understanding the
 vacuum selection effect is the effective separation between the
 dynamics of the magnetic core and of the modulus field outside the
 core (this separation is only true for vortices, as it is a
 consequence of the properties of massless fields in two dimensions,
 the transverse dimensions of the vortex).  Far from the core the
 magnetic field vanishes, there is no potential energy, and the
 modulus field (properly normalised) becomes massless, with solutions
 logarithmic in $r$ -- unless $u=0$.  Thus $u=0$ is selected.

We can see this in another way:
well outside the core the scalar fields can move along 
the moduli space from $u=0$ to their asymptotic values, {\it with no 
appreciable cost in energy}. Indeed, outside the core,
\[
E \sim \int d^2x\, (D_\mu \phi_+)^2 + (D_\mu \phi_-)^2 \sim
\int d^2x (\partial_r u)^2 \cosh 2u 
\]
and the minimum energy configuration $u(r)$ that interpolates between
any two distinct vacua, $u(R_1)=u_1 $ and $u(R_2) = u_2$, if $R_1, R_2 >> 
r_{\text core}$, is of the form
\[
 E \sim {I(u_1, u_2) \over {\ln R_2 - \ln R_1}},
\hspace{0.5cm}
{I (u_1, u_2)\over{2\pi}} =\left[
 \int_{u_1}^{u_2}\!\! du\sqrt{\cosh 2u}\right]^2
\]
which can be made arbitrarily small as $R_2 \to \infty$.  Thus, the
dynamics at the core are decoupled from the behaviour of the moduli
fields; the core of the vortex is effectively free to choose its
boundary conditions in order to minimise its own energy.  At the core
(where the magnetic field is non-zero) the field $\phi_-$ is
suppressed and the core becomes identical to that of a NO string.  But
the previous argument also shows that the minimum energy is
unattainable unless $I=0$ ($u_1=u_2$), and this is a condition for
static solutions.  Thus, the vacuum selection actually occurs {\it at
the core}.  

The authors of \cite{PRTT96} offered no insight into why a particular
vacuum is selected, but it is clear that $\phi_-=0$ is chosen because
it minimises the mass of the gauge field, $m_v$.  This is an important
consideration at the core of~the vortex, where the magnetic field is
concentrated in a region of order $m_v^{-1}$. Magnetic field lines
repel, so a lower $m_v$ means a larger magnetic core. Since the total
magnetic flux is quantised, this lowers the energy at the core and,
therefore, the total energy.

\section{BPS vortices in $N\!=\!2$ SQED and vacuum selection}

We now consider two N=2 hypermultiplets $h_a, \ a=1,2$ with charges
$q_a = +q,-q$ coupled to an N=2 Abelian vector multiplet \cite{ADH98}.
This model has been analysed before as a toy model for the low energy
effective action of Calabi--Yau compactifications of type II
superstrings by Greene, Morrison and Vafa\cite{GMV96}, who identified
magnetic configurations due to the wrapping of D--branes on cycles.
In the low energy theory these would appear as monopole--antimonopole
pairs joined by non--topological magnetic vortices providing magnetic
confinement.

In \cite{ADH98} it was shown that, in the absence of FI terms, such
vortices were unstable. In an attempt to stabilise the vortices, here
we consider the addition of FI D-terms, the only modification still
compatible with N=2 supersymmetry of the effective action.  We will be
partially successful in that the instability turns into a zero mode,
but as we will show this is still not enough to provide confinement.

The model has a global SU(2) symmetry that rotates the two scalar
fields in each multiplet, and we introduce a triplet of FI D-terms
$\vec{k}\cdot \vec{D}$. In this case the answer can be more involved:
since the vacuum manifold is simply connected there is no unique
vacuum that minimises the vector mass and it is not clear that the
selected vacua can lead to vortices.  We take $\vec k = (0,0,\frac12
|q|\omega^2)$, without loss of generality; note that this breaks the
SU(2) symmetry.

Eliminating the auxiliary fields and setting to zero 
those fields that have zero vacuum expectation values, simplifies the
model considerably. With a suitable rescaling
$h_{ij}\to \omega h_{ij}$, $x_{\mu}\to {x_{\mu}}/{\omega q}$, 
$A_{\mu}\to\omega A_{\mu}$, 
the energy of static, straight vortex configurations can be written as
\begin{eqnarray}
{{2E}\over{\omega^2}} & = & \int\!\! d^2x\left[|D_\mu h_{11}|^2 +
     |D_\mu h_{12}|^2 + |D_\mu h_{21}|^2  +|D_\mu h_{22}|^2 +\frac12 B^2 \right. \nonumber \\
     & &\left. 
       +(H^{\;1}_1-H^{\;2}_2 + 1/2)^2
         + (H^{\;1}_2 + H^{\;2}_1)^2+ (i\,H^{\;1}_2 - i\,H^{\;2}_1)^2 \right] 
\label{model}
\end{eqnarray}

where $D_{\mu} = \partial_\mu + {i (q_a/q)} A_\mu$ and 
$H^{\;\;i}_j = - ({q_a}/{2q})\,h_{ai}^*h_{aj} $.

There is another $SU(2)$ symmetry between ($h_{11}$ and $h_{22}^*$) 
and ($h_{12}$ and $h_{21}^*$) which is preserved after adding the D-term.
For finite energy we require $D_\mu h_{ai} = 0 $ as $r \to \infty$, 
which correlates the phases of the multiplets at spatial infinity, 
giving $h_a \sim e^{inq_a \theta}$ and the quantisation of magnetic flux
from $A_\theta d\theta\sim -n d\theta$; 
the scalars must lie in the vacuum manifold 
($H^{\;1}_2 = H^{\;2}_1 = 0$, $H^{\;2}_2-H^{\;1}_1 =\frac12$). 

Parametrising the scalars  as $
 h_{ai} = r_{ai}\, e^{\,i\,\chi_{ai}}$, gives
\begin{eqnarray} 
 & &\chi_{11}-\chi_{12}=\chi_{21}-\chi_{22}+2m\pi\,,\\
& &r_{11}\,r_{12} - r_{21}\,r_{22} =  0 \,,\label{eq:r2} \\ 
 & &(r_{11})^2 + (r_{22})^2 \, - (r_{12})^2 - (r_{21})^2  =  1\,.\label{eq:r4}
\end{eqnarray}
The vacuum manifold is simply 
connected, but the set of finite energy configurations has 
non-contractible loops labelled by $n$,
the (quantised) magnetic flux. This situation is familiar in the context
of semilocal models (see \cite{AV00} for a recent review), where
the gauge field couples to two (or more) scalar fields of {\it equal} charges,
and there is a global SU(2) (or larger) symmetry between the scalars.
The stability of semilocal vortices depends on the masses 
of the scalar and
the vector particles \cite{H92}. If $m_s < m_v$ the strings are stable. 
In the Bogomol'nyi limit $m_s=m_v$, 
the resulting  $n=1$ ``vortices'' have a richer structure including a 
scalar condensate at the core which causes the magnetic flux 
to spread  over an arbitrarily large area. The thinnest vortex 
is a NO string but it is not protected against 
the zero mode that generates the  scalar condensate; as a result its 
energy density becomes arbitrarily close to the vacuum in a finite time
and the string dissolves \cite{leese}.
If $m_s>m_v$ the zero mode becomes an instability and there are no strings.

Let us consider the vacuum selection effect in model (\ref{model}).
The mass of the gauge field is $r_{11}^2 + r_{12}^2 + 
r_{21}^2 + r_{22}^2 $, and the minimization of this mass, subject to eqns (\ref{eq:r2},\ref{eq:r4}), 
predicts $h_{12} = h_{21} = 0$. 
To prove that this is the case, we rewrite the energy as 
\begin{eqnarray}
2E&=& \int d^2x\left[ 
|(D_1+iD_2) h_{11}|^2 \ + \  |(D_1-iD_2)h_{12}|^2
\right. \nonumber\\
& &+  |(D_1+iD_2) h_{21}|^2 \ + \ |(D_1-iD_2)h_{22}|^2
\nonumber\\
& &+  [ B +  (H^{\;\;1}_1 - H^{\;\;2}_2 + \frac12)]^2 + \,(H^{\;\;1}_2 
+ H^{\;\;2}_1)^2\nonumber\\
& &+\left.  \, (i\,H^{\;\;1}_2 - i\,H^{\;\;2}_1)^2\right] - \int d^2x\, B \,,
\end{eqnarray}
and obtain the Bogomol'nyi equations:
\begin{eqnarray}
&(D_1+iD_2) h_{11} = 0 \qquad (D_1-iD_2)h_{22} = 0& \label{22}\\
&(D_1-iD_2)h_{12}  = 0 \qquad (D_1+iD_2) h_{21} = 0& \label{21} \\
&H^{\;\;2}_1 =  0 &\label{H21}\\
&B +  \left[H^{\;\;1}_1 - H^{\;\;2}_2 + 1/2\right] = 0 & \ . \label{Beq} 
\end{eqnarray}

(Note that  setting 
$ H^{\;\;2}_1 =  0 $ implies
$D_+  H^{\;\;2}_1 = 0$ which holds if 
$D_+ h_{11}    = D_+ h_{12}^* =   D_+ h_{21} =  D_+ h_{22}^* = 0$ 
so the choice of $\pm$ signs in the gradient terms is not arbitrary).

The most
general configuration for the scalar hypermultiplets compatible
with cylindrical symmetry is \cite{ADH98}
\begin{eqnarray}
&h_1  \equiv  \pmatrix{h_{11} (r) \cr h_{12} (r) \cr} e^{i\theta},\qquad 
h_2  \equiv  \pmatrix{h_{21} (r) \cr h_{22} (r) \cr}
 e^{\,i\,\Delta} e^{-i \theta}\,, &
 \nonumber \\
&A_{\theta} = a(r)  \qquad  A_r =  0&\,.
\label{cylsym}
\end{eqnarray}

To analyse the Bogomol'nyi equations we simply note that the pairs of
fields $(h_{11} , h_{21})$ and $(h_{22}^*, h_{12}^*)$ behave exactly like
$(\phi_+, \phi_-)$ in the N=1 model of the previous section: 
they have charges $\pm1$ and they appear in the square bracket of 
eq. (\ref{Beq}) with 
signs +1, -1 respectively. The only way to satisfy (\ref{21}) 
is to have $h_{12}=h_{21}=0$, which satisfies (\ref{H21}) automatically.
These configurations minimise the vector mass, so 
we conclude that the ``vacuum selection'' effect  also works
here (note that by setting the second hypermultiplet to zero
the vacuum selection effect gives a topological vortex in $h_{11}$
\cite{hou}).

This leaves eqns. (\ref{22}), (\ref{Beq}), 
which are precisely the Bogomol'nyi equations 
of semilocal strings \cite{VA91} in $(h_{11}, h_{22}^*)$. 
These have been studied in \cite{H92,GORS92}, so we simply state the results. 
For $n=1$, the non-zero components of the most general cylindrically 
symmetric solution are 
\[
(h_{11}\,,\,h_{22}^*) =
\left(f(r)e^{i\theta}\,,\,p(r)e^{i\Delta}\right)U
\]
where $U$ is an arbitrary global $SU(2)$ rotation,
$\Delta$ is a constant and
$f(r)$ and $p(r)$ are real functions with  
$f(0) = 0$, $f(\infty) = 1$ and $p(0) \neq 0$, $p(\infty) = 0$. They 
are related by $p(r) = \alpha f(r) /r$, where $\alpha$ is a real positive
parameter which labels the solutions. For each $\alpha$
the gauge field is obtained from the condition $(D_1+iD_2 )h_{11} = 0$.

The lack of winding in $p$ results in the scalar condensate at 
the core, $p(0) \neq 0$. Note also that the 
SU(2) transformations between $h_{11}$ and $h_{22}^*$
are not related to the SU(2) transformations 
coming from the N=2 supersymmetry, since they mix elements 
of the two hypermultiplets. The zero mode is not related to {\textbf either} 
of these global symmetries, as can be seen by the fact that it does not
alter the boundary conditions at infinity. 
To understand its effect, we
consider the asymptotic behaviour of the fields far from the core
\cite{leese,m93}:
\begin{eqnarray}
f(r) &\sim& 1 - \frac12 (\alpha^2/r^2) + \alpha^2(3/8 \alpha^2-2)/r^4 +\dots \nonumber\\
a(r) &\sim& -1 + (\alpha^2/r^2) - \alpha^2(\alpha^2 -8)/r^4+ \dots
\end{eqnarray}

There is a one-parameter set of magnetic ``vortices''
with varying widths.
They all saturate the Bogomol'nyi bound, so they are all degenerate in energy,
but they have different structures: most notably, they have a 
scalar condensate at the core and the magnetic field fall--off 
is a power of $r$ (in NO vortices the fall-off is exponential).
These ``vortices'' can be thought of as hybrids between the NO vortex 
and the $CP^1$ lumps, and
$\alpha$ can be seen 
as the width above that 
of a NO vortex: $\alpha = 0$ corresponds to the NO string modulo SU(2)
 rotations; but as $\alpha \to \infty$ $p(0) \to 1$ and the 
core expands.

Finally, the zero mode corresponds to a flat direction 
in the potential for which the vector mass is 
always at a minimum, so the vacuum selection effect 
does not single out the NO solution over the more extended vortices.

We conclude that magnetic flux in this model is still not confined to
tubes of a definite size. Semilocal strings which saturate the
Bogomol'nyi bound are at the limiting case between stability (for $m_s
< m_v$) and instability (for $m_s> m_v$) \cite{H92,AV00}.  However,
\cite{leese} has shown that, in fact, the flux is unconfined and
eventually spreads out to flux tubes of greater and greater radius. Of
crucial importance to the resulting cosmology is the time-scale for
this relaxation, which is outside the scope of this investigation, but
it is clear that vorton production will be suppressed.

Our conclusions are easily extended to any number of hypermultiplets,
as long as their charges are equal in absolute value. They also apply
in the case of several gauge fields, such as in the low energy
limit of type II superstrings compactified on Calabi--Yau manifolds,
which effectively contains 15 copies of the model analysed here
\cite{GMV96}.

The situation changes when considering hypermultiplets with different
charges. In that case vacuum selection may be frustrated and the
selected vacuum or vacua may not be the ones that minimise the vector
mass. The structure of the resulting vortices can be totally different
from what we described here, with the possibility of binary or
multiple cores and other exotic effects and it remains a very
interesting open question.

\section{Discussion}

We have investigated cosmic string solutions in SUSY QED
 with flat directions where the U(1) symmetry is broken by a
FI D-term.  We have argued that, if all
matter fields have the same absolute value of the charge, the vacuum
selection effect acts to minimise the vector
boson mass.  This includes the original model of
\cite{PRTT96}, the bosonic sector of N=1 SUSY QED with two
chiral superfields of opposite charges, and its simplest extension to
N=2 with one hypermultiplet, considered in \cite{hou}. In both cases
the vacuum selection effect leads to topological vortices; we have
shown that both solutions saturate Bogomol'nyi bounds and are
consequently stable.

We then considered extensions to N=2 SUSY QED with several
hypermultiplets of equal $|q_a|$. The case of two hypermultiplets of
opposite charge was solved explicitly, and there is still a vacuum selection
effect, but this case gives rise to a semilocal string  instead. Again
the string solution saturates the Bogomol'nyi bound, but is only 
critically stable as there is a zero mode (which is {\textit {not}} an SU(2) rotation). 
The solutions that saturate 
the Bogomol'nyi bound are parametrised by one or more internal 
degrees of freedom and are distinct from the string solution, 
in particular their core widths can be arbitrarily large. 
Note that our $N=2$ results generalise to Abelian theories with $M$  
hypermultiplets provided they have equal $|q_a|$; one would still find
 semilocal strings, but the internal symmetry would be $SU(M)$ 
and not $SU(2)$ \cite{m93}. Magnetic flux would still be unconfined. 

The situation changes when considering hypermultiplets with different charges. In that case vacuum selection may be frustrated and the selected vacuum os vacua may not be the ones that minimise the vector mass. The structure of the resulting vortices can be totally different from what we describe here, with the possibility of binary or multiple cores and other exotic effects and it remains a very interesting open question. 

As a consequence of supersymmetry, there will be fermion zero
modes in the string core for both models.  The zero modes can be found
using the results of \cite{DDT97}.  For the model of \cite{PRTT96}
there will be a single zero mode, left or right moving depending on
the sign of $n$, which is a combination of the Higgsino and
gaugino. Thus the string will be chiral \cite{chiralstring}, since it
has a chiral current. We can then apply the results of
\cite{chiralvorton} to this model.  If there are no further phase
transitions to destabilise the current, the model will lead to vorton
formation, restricting the scale of symmetry breaking
\cite{chiralvorton}.  As shown here, this is in contrast with the
situation in the $N\!=\!2$ model, where vorton production is not expected
to occur.

However, if we completely break SUSY, for instance via mass terms for the $h_{ij}$ fields, we still expect a topological string with fermion zero modes in the core. The angular momentum of the resulting current can stabilise loops, giving rise to vorton production \cite{ADPU02}. The
model will then be subject to the constraints of \cite{chiralvorton}.

\section{Acknowledgements}
This work is supported in part by PPARC, CICYT AEN99-0315 grant 
and the ESF COSLAB programme. JU is partially supported by a Marie Curie Fund 
of the EC programme HUMAN POTENTIAL HPMT-CT-2000-00096.
JU thanks DAMTP, ACD thanks the UPV  for hospitality. 
We thank the Basque Government and the British Council 
for a visiting grant. AA thanks E. Jim\'enez for 
invaluable help. We thank M. de Roo, M. Bianco and J.M. Evans
for discussions.

\end{document}